\documentclass[a4paper,11pt]{article}
\usepackage[margin=1in]{geometry}
\usepackage{cite}
\usepackage{amsmath,amssymb,amsfonts}
\usepackage{algorithmic}
\usepackage{graphicx}
\usepackage{textcomp}
\usepackage{xcolor}
\usepackage{amsthm}
\usepackage{enumitem,hyperref}

\newtheorem{theorem}{Theorem}
\newtheorem{lemma}{Lemma}
\newtheorem{corollary}{Corollary}

\newtheorem{conjecture}{Conjecture}

\newcommand{\E}{\mathbb{E}}
\newcommand{\R}{\mathbb{R}}

\newcommand{\Var}{\mathrm{Var}}
\newcommand{\Ent}{\mathrm{Ent}}
\newcommand{\BSC}{\mathrm{BSC}}
\newcommand{\prob}{\mathbb{P}}

\def\BibTeX{{\rm B\kern-.05em{\sc i\kern-.025em b}\kern-.08em
    T\kern-.1667em\lower.7ex\hbox{E}\kern-.125emX}}
\begin{document}

\title{Progress on the Courtade-Kumar Conjecture: Optimal High-Noise Entropy Bounds and Generalized Coordinate-wise Mutual Information}

% \author{\IEEEauthorblockN{Adel Javanmard}
% \IEEEauthorblockA{\textit{Department of Data Sciences and Operations} \\
% \textit{University of Southern California and Google Research}\\
% Los Angeles, CA, USA \\
% ajavanma@usc.edu}
% \and
% \IEEEauthorblockN{David P. Woodruff}
% \IEEEauthorblockA{\textit{Department of Computer Science} \\
% \textit{Carnegie Mellon University and Google Research}\\
% Pittsburgh, PA, USA \\
% dwoodruf@cs.cmu.edu}
% }
% \author[1,3]{Adel Javanmard}
% \author[2,3]{David P. Woodruff}
% \affil[1]{Department of Data Sciences and Operations
% University of Southern California, Los Angeles, CA, USA \texttt{ajavanma@usc.edu}}
% \affil[2]{Department of Computer Science
% Carnegie Mellon University, Pittsburgh, PA, USA \texttt{dwoodruf@cs.cmu.edu}}
% \affil[3]{Google Research}

\author{
    Adel Javanmard\thanks{Department of Data Sciences and Operations,
University of Southern California, Los Angeles, CA, USA \texttt{ajavanma@usc.edu}}, \thanks{Google Research} \and 
    David P. Woodruff\thanks{Department of Computer Science,
Carnegie Mellon University, Pittsburgh, PA, USA, \texttt{dwoodruf@cs.cmu.edu}}, \footnotemark[2]
}

%\date{}

\maketitle

\begin{abstract}
The Courtade-Kumar conjecture~\cite{courtade2014boolean} posits that dictatorship functions maximize the mutual information between the function's output and a noisy version of its input over the Boolean hypercube. We present two significant advancements related to this conjecture. First, we resolve an open question posed by Courtade and Kumar~\cite{courtade2014boolean}, proving that for any Boolean function (regardless of bias), the sum of mutual information between the function's output and the individual noisy input coordinates is bounded by $1-H(\alpha)$, where $\alpha$ is the noise parameter of the Binary Symmetric Channel. This generalizes their previous result which was restricted to balanced Boolean functions. Second, we advance the study of the main conjecture in the high noise regime. We establish an optimal error bound of $O(\lambda^2)$ for the asymptotic entropy expansion, where $\lambda = (1-2\alpha)^2$, improving upon the previous best-known bounds. This refined analysis leads to a sharp, linear Fourier concentration bound for highly informative functions and significantly extends the range of the noise parameter $\lambda$ for which the conjecture is proven to hold.
\end{abstract}

%\begin{IEEEkeywords}


\section{Introduction}

The study of Boolean functions under noise is a cornerstone of theoretical computer science and information theory, with deep connections to coding theory, hardness of approximation, and learning theory. A fundamental question in this area concerns the resilience of information transmission when the input to a function is corrupted by noise.

Consider the Boolean hypercube $\{-1, 1\}^n$. Let $X$ be drawn uniformly from this space. Suppose $X$ is transmitted through a memoryless Binary Symmetric Channel with crossover probability $\alpha \in [0, 1/2]$, denoted $\BSC(\alpha)$, resulting in the noisy output $Y$. We are interested in the mutual information $I(b(X); Y)$ for a Boolean function $b: \{-1, 1\}^n \to \{-1, 1\}$.

In their influential work, Courtade and Kumar \cite{courtade2014boolean} conjectured that this mutual information is maximized by the simplest functions, namely the dictatorships (e.g., $b(x) = x_1$).

\begin{conjecture}[Courtade-Kumar Conjecture \cite{courtade2014boolean}]\label{conj:CK}
For any Boolean function $b : \{-1, 1\}^n \to \{-1, 1\}$,
$$ I(b(X); Y) \leq 1 - H(\alpha). $$
\end{conjecture}

The quantity $1-H(\alpha)$ is the capacity of the $\BSC(\alpha)$. The conjecture suggests that concentrating the function's dependence on a single coordinate is the optimal strategy for preserving information through noise.

Despite significant interest, Conjecture \ref{conj:CK} remains open. Progress has been made in specific regimes. Notably, Samorodnitsky \cite{samorodnitsky2016entropy} proved the conjecture in the high noise regime, i.e., when $\alpha$ is sufficiently close to $1/2$. This regime corresponds to the noise correlation parameter $\rho = 1-2\alpha$ being close to 0.

In addition to their main conjecture, Courtade and Kumar proved a related statement concerning the sum of mutual information with individual coordinates.

\begin{theorem}[Theorem 1 in \cite{courtade2014boolean}]\label{thm:CK_Thm1}
If $b$ is a balanced Boolean function ($\E[b]=0$), 
$$ \sum_{i=1}^n I(b(X); Y_i) \leq 1 - H(\alpha). $$
\end{theorem}

They posed as Open Question 1 whether this result extends to arbitrary, potentially biased, Boolean functions.

Beyond the primary Courtade-Kumar conjecture, several related conjectures have been proposed in \cite{li2020boolean, anantharam2017conjecture} (see also~\cite{barnes2020courtade}). These works generally posit that dictatorship functions are the maximizers for various functionals of Boolean functions.

\subsection{Our Contributions}

This paper provides two main contributions towards understanding the information-theoretic properties of Boolean functions under noise.

\subsubsection{Resolution of Open Question 1}
We provide a complete and affirmative resolution to the open question posed by Courtade and Kumar. We prove that Theorem \ref{thm:CK_Thm1} holds for all Boolean functions.

\begin{theorem}[Generalized Coordinate-wise Bound]\label{thm:generalized_thm1_intro}
For any Boolean function $b : \{-1, 1\}^n \to \{-1, 1\}$,
$$ \sum_{i=1}^n I(b(X); Y_i) \leq 1 - H(\alpha). $$
\end{theorem}

The proof, presented in Section \ref{sec:OQ1}, involves analyzing the optimization landscape. We show that the objective function, viewed as a function of the squared level-1 Fourier coefficients, is convex. We analyze the extreme points of the feasible region defined by Parseval's identity and bounds on Fourier coefficients. The proof requires establishing a non-trivial inequality involving derivatives of the mutual information expression with respect to $\rho$. A key distinction from the proof strategy in~\cite[Theorem 1]{courtade2014boolean} is our demonstration that the objective function is maximized by Boolean functions that are monotone in each coordinate (defined via a 1-d compression operator). Using this structure we can impose tighter constraints on the feasible set which allows us to generalize the result of~\cite[Theorem 1]{courtade2014boolean} to all Boolean functions.

\subsubsection{Optimal Entropy Bounds in High Noise}
We significantly refine the analysis of Conjecture \ref{conj:CK} in the high noise regime. We introduce the noise parameter $\lambda = \rho^2 = (1-2\alpha)^2$, which is small in this regime. The approach initiated by \cite{samorodnitsky2016entropy} relies on an asymptotic expansion of the entropy of the noisy function, $\Ent(T_\alpha f)$, where $T_\alpha$ is the noise operator corresponding to $\BSC(\alpha)$, defined as
defined as:
\begin{equation}
    T_\alpha f(x) = E(f(Y) \mid X = x)\,,
\end{equation}
where $f$ is defined on the hypercube $\{-1,1\}^n$.

Note that $T_\alpha$ is a linear mapping and can be expressed as
$$(T_\alpha f)(x) = \sum_{y \in \{0,1\}^n} \alpha^{d(x,y)} (1-\alpha)^{n-d(x,y)} f(y),$$
where $d(\cdot,\cdot)$ is the Hamming distance. Previous work established the following bound for any nonnegative non-zero function $f$ (normalized without loss of generality to satisfy $\E[f] =1$):
$$ \Ent(T_\alpha f) \leq \left(\frac{L_1(f)}{2 \ln 2}\right) \cdot \lambda + E(\lambda), $$
where $L_1(f)$ is the level-1 Fourier energy, given by $L_1(f)= \sum_{i\in[n]}\hat{f}(\{i\})^2$, and $E(\lambda) = O(\lambda^{4/3})$~\cite{samorodnitsky2016entropy}. Our next result improves this bound to $O(\lambda^2)$ and proves its optimality (recall that $\lambda\in [0,1]$.) 
 While the rate $O(\lambda^2)$ is anticipated to be the correct asymptotic behavior (as it matches the expansion for the dictatorship function), rigorously establishing this upper bound for general functions requires delicate handling of the higher-order terms.

\begin{theorem}[Optimal Asymptotic Entropy Bound]\label{thm:optimal_entropy_intro}
For any bounded non-negative function $f$ with $\E[f] = 1$:
$$ \Ent(T_\alpha f) \leq \left(\frac{L_1(f)}{2 \ln 2}\right) \cdot \lambda + O(\lambda^2). $$
\end{theorem}

The proof, detailed in Section \ref{sec:high_noise}, relies on a direct Taylor expansion approach combined with rigorous moment bounds derived using hypercontractivity. Crucially, we employ a careful application of the Minkowski inequality over the Fourier decomposition to establish the necessary higher-moment bounds. The optimality of this bound is demonstrated by the fact that the $O(\lambda^2)$ error rate is tight for Boolean dictatorship functions. 
%Due to space constraint, we defer the proof of Theorem~\ref{thm:optimal_entropy_intro} to the supplementary material~\cite{?}.

This optimal error bound has profound implications for the structure of highly informative functions.

\begin{theorem}[Linear Fourier Concentration]\label{thm:linear_concentration_intro}
If a Boolean function $f$ (with bounded bias) achieves near-maximal mutual information, $I(f(X); Y) \geq 1 - H(\alpha)$, then
$$ \sum_{|S|\geq 2} \hat{f}(S)^2 = O(\lambda). $$
\end{theorem}

This linear concentration bound (improving the prior bound of $O(\lambda^{1/3})$ from \cite{samorodnitsky2016entropy}) is the tightest possible within this framework. It confirms that functions maximizing information must concentrate their Fourier mass on level 1.

Consequently, this allows us to significantly extend the range of the noise parameter $\lambda$ for which the Courtade-Kumar conjecture is proven to hold. We show there exists a threshold $\delta_{opt} > 0$ such that the conjecture holds for all $\lambda < \delta_{opt}$, and this threshold is strictly larger than that established in \cite{samorodnitsky2016entropy}. We formalize it in the next theorem.
\begin{theorem}[Extended Range for the Conjecture]\label{thm:extended_range}
There exists an absolute constant $\delta_{opt} > 0$ such that if the noise parameter $\lambda \leq \delta_{opt}$, the Courtade-Kumar conjecture holds. Furthermore, $\delta_{opt}$ is strictly larger than the threshold established using the $O(\lambda^{4/3})$ error bound in \cite{samorodnitsky2016entropy}.
\end{theorem}

\subsection{Organization}
The paper is organized as follows. Section \ref{sec:OQ1} presents the proof of Theorem \ref{thm:generalized_thm1_intro}. Section \ref{sec:high_noise} introduces the setup for the high noise regime and derives Theorem \ref{thm:extended_range}, and its subsections provides the proof of Theorems~\ref{thm:optimal_entropy_intro} and Theorem~\ref{thm:linear_concentration_intro}. In Section~\ref{sec:applications} we discuss several potential applications of our results.

\section{Generalized Coordinate-wise Mutual Information Bound}\label{sec:OQ1}

In this section, we prove Theorem \ref{thm:generalized_thm1_intro}, resolving Open Question 1 of \cite{courtade2014boolean}. 
%Due to space constraint, proof of lemmas are deferred to the supplementary material~\cite{?}. 

% \begin{theorem}[Generalized Theorem 1]
% Let $X^n$ be i.i.d. uniform on $\{-1, 1\}^n$, and let $Y^n$ be the output of a $\BSC(\alpha)$ with input $X^n$. Let $\rho = 1 - 2\alpha$. For any Boolean function $b : \{-1, 1\}^n \to \{-1, 1\}$,
% $$ \sum_{i=1}^n I(b(X^n); Y_i) \leq 1 - H(\alpha). $$
% \end{theorem}

% \begin{proof}
\subsection{Proof of Theorem~\ref{thm:generalized_thm1_intro}}
Recall the crossover probability $0<\alpha<\frac{1}{2}$ and define $\rho=1-2\alpha$. Any Boolean function $b:\{-1,+1\}^n\to \{-1,1\}$ can be written in terms of its Fourier coefficients as
\[
b(x) = \sum_{S\subseteq[n]}\hat{b}(S)\Pi_S(x)\,,
\]
where $\Pi_S(x) = \Pi_{i\in S} x_i$ are the orthonormal basis for the Fourier transform and $\{\hat{b}(S)\}_{S\subseteq[n]}$ are the Fourier coefficient defined by
\[
\hat{b}(S)= \E b(X)\Pi_S(X)\,.
\]
For $S=\emptyset$, we define $\Pi_S(x^n)= 1$.

Let $\mu=\E[b]=\hat{b}(\emptyset)$ be the bias, and $z_{i}=\hat{b}(\{i\})$ be the level-1 Fourier coefficients. The objective function is:
\begin{align}\label{eq:Lb}
L(b)=\sum_{i=1}^{n}I(b;Y_{i})=\sum_{i=1}^{n}(H(b)-H(b|Y_{i})).
\end{align}
Since the range of $b$ is $\{-1,+1\}$, the entropy $H(b)$ is determined by the bias $\prob(b=1) = \frac{1+\E[b]}{2} = \frac{1+\mu}{2}$, hence we can write $H(b)=H(\frac{1+\mu}{2})$. Using  relation (34) from~\cite{courtade2014boolean}, we have
\[
\prob(b=1|Y_i=y_i) = \frac{1+\mu+\rho y_i z_i}{2}\,,
\]
and since the marginal distribution of $Y_i$ is uniform on $\{-1,+1\}$, we get
\begin{align}
H(b|Y_i) = \frac{1}{2} \left[ H(\prob(b=1|Y_i=1))+H(\prob(b=1|Y_i=-1))
\right]\,.
\end{align}
Therefore we obtain $H(b|Y_{i})=h_{\mu}(z_{i}),$ where
\[
h_{\mu}(z):=\frac{1}{2}H\left(\frac{1+\mu+\rho z}{2}\right)+\frac{1}{2}H\left(\frac{1+\mu-\rho z}{2}\right).
\]
Let $g_{\mu}(z)=H(b)-h_{\mu}(z)$. By~\eqref{eq:Lb}, we want to maximize $\sum_{i=1}^{n}g_{\mu}(z_{i})$ subject to the constraints on the Fourier coefficients of a Boolean function.

\bigskip
\noindent\textbf{Step 1: Reduction to Monotone Functions.}
Let $f: \{0, 1\}^n \to \{0, 1\}$ be a Boolean function. The 1-d compression operator along coordinate $i$, denoted as $\mathcal{C}_i$, rearranges the values of the function along the $i$-th dimension to make it ``monotone''  while preserving the total number of 1s. Specifically, for any input $x \in \{0, 1\}^n$, let $x = (x_{\sim i}, x_i)$, where $x_{\sim i}$ are all bits except $i$. The operator acts on the pair of values $(f(x_{\sim i}, 0), f(x_{\sim i}, 1))$:

\begin{equation}
\mathcal{C}_i f(x_{\sim i}, x_i) =
\begin{cases}
    1 & \text{if } x_i = 1 \text{ and } f(x_{\sim i}, 0) + f(x_{\sim i}, 1) \ge 1 \\
    1 & \text{if } x_i = 0 \text{ and } f(x_{\sim i}, 0) + f(x_{\sim i}, 1) = 2 \\
    0 & \text{otherwise}
\end{cases}
\end{equation}
We show that 1-d compression increases
the objective function $L(b)$. 
\begin{lemma}\label{lem:fourier-prop}
Let $\tilde{b}$ be the compression of $b$ along coordinate $j$, making it non-decreasing in $x_j$. The Fourier coefficients satisfy:
\begin{enumerate}
\item $z_i(\tilde{b}) = z_i(b)$ for $i\neq j$.
\item $|z_j (\tilde{b})| \ge |z_j (b)|$.
\end{enumerate}
\end{lemma}
Note that $h_\mu(z) =  h_\mu(-z)$ and the function $h_{\mu}(z)$ is decreasing in $|z|$, hence by Lemma~\ref{lem:fourier-prop} $h_{\mu}(z_{j}(\tilde{b}))\le h_{\mu}(z_{j}(b))$.

Thus, $\sum_i H(b|Y_i)$ decreases under compression
and $L(b)=\sum_i I(b; Y_i)$ increases. This implies that the maximum must be attained by a function $b$ that is monotone along each coordinate.
We assume $b$ is non-decreasing, so $z_i(b) \ge 0$, for $i\in[n]$, by the next lemma.
\begin{lemma}\label{lem:xi-pos}
Suppose that $b$ is a non-decreasing Boolean function, i.e., for  any input $x$, changing the $i$-th bit from $-1$ to $1$ either increases the function value or keeps it the same: $b(x_{\sim i}, -1)\le b(x_{\sim i}, +1)$. Then $z_i(b):=\hat{b}(\{i\}) \ge 0$.
\end{lemma}

We will also use the following lemma in the next step.

\begin{lemma}\label{lem:boolean-prop}
For any Boolean function $b:\{-1,1\}^n\to\{-1,1\}$, with
$\mu = \E[b]$ and $z_i = \hat{b}(\{i\})$ the level-1 Fourier coefficients, we have $z_i \le 1-|\mu|$.
\end{lemma}

\bigskip
\noindent\textbf{Step 2: Optimization over Monotone Functions.}
We analyze the structure of the optimization problem. Let $w_{i}=z_{i}^{2}$. Define $\psi_{\mu}(w)=g_{\mu}(\sqrt{w})$. It can be shown that $\psi_{\mu}(w)$ is strictly convex for $w>0$ (Equivalently, $h_{\mu}(\sqrt{w})$ is strictly concave in $w$).

We want to maximize the convex function $\sum_{i=1}^{n}\psi_{\mu}(w_{i})$ subject to constraints on the Fourier coefficients. 

The constraints are:
\begin{enumerate}
    \item $w_{i}\ge0$
    \item $\sum_{i=1}^{n}w_{i}\le1-\mu^{2}=R^{2}$ (Parseval's theorem).
    \item $w_{i}\le(1-|\mu|)^{2}=C^{2}$ (Since $0\le z_{i}\le1-|\mu|$ by Lemmas~\ref{lem:xi-pos} and \ref{lem:boolean-prop}).
\end{enumerate}
The maximum of a convex function over this polytope is attained at an extreme point. Let $K=R^{2}/C^{2}=(1-\mu^{2})/(1-|\mu|)^{2}=(1+|\mu|)/(1-|\mu|)$. We analyze two cases based on the relationship between $n$ and $K$. Without loss of generality, assume $\mu\ge0$. 

\bigskip
\noindent\textbf{Step 3: Bounding the Maximum (Case 1: $n\ge K$).}
If $n\ge K,$ the constraint $\sum w_{i}\le R^{2}$ is the dominant constraint. The extreme points (up to permutation) have $k=\lfloor K\rfloor$ coordinates equal to $C^{2}$, one coordinate equal to $\theta C^{2}$ (where $\theta=K-k)$, and the rest are $0$. The maximum value is:
\[
M(\mu)=k\psi_{\mu}(C^{2})+\psi_{\mu}(\theta C^{2}).
\]
We verify that $\psi_{\mu}(0)=0.$ Note that $\psi_{\mu}(0)=g_{\mu}(0)=H(b)-h_{\mu}(0)$. Since $h_{\mu}(0)= H(\frac{1+\mu}{2})=H(b)$, we have $\psi_{\mu}(0)=0$. Using the convexity of $\psi_{\mu}(w)$ and $\psi_{\mu}(0)=0$:
\begin{align*}
\psi_{\mu}(\theta C^{2})&=\psi_{\mu}(\theta C^{2}+(1-\theta)0)\\
&\le\theta\psi_{\mu}(C^{2})+(1-\theta)\psi_{\mu}(0)=\theta\psi_{\mu}(C^{2}).
\end{align*}
Thus, the maximum is bounded by:
\begin{align*}
M(\mu)&\le k\psi_{\mu}(C^{2})+\theta\psi_{\mu}(C^{2})\\
&=(k+\theta)\psi_{\mu}(C^{2})\\
&=K\psi_{\mu}(C^{2})=Kg_{\mu}(C).
\end{align*}
Recall that $C = 1-\mu$ and $K = (1+\mu)/(1-\mu)$, and so $\mu$ and $C$ are determined by $K$ as follows:
\[
\mu = \frac{K-1}{K+1}\,, \quad C = \frac{2}{K+1}\,.
\]
Let $M_{K}(\rho):= Kg_{\mu}(C),$ where we made the dependence on $\rho = 1-2\alpha$ explicit in the notation. We aim to show $M_{K}(\rho)\le1-H(\alpha)$.

Writing $M_K(\rho)$ more explicitly, we have
\begin{align*}
&M_K(\rho)\\
&= K H(b) - \frac{K}{2}H\left(\frac{1+\mu+\rho C}{2}\right)-\frac{K}{2}H\left(\frac{1+\mu-\rho C}{2}\right)\\
&= K H\left(\frac{K}{K+1}\right)-\frac{K}{2}H\left(\frac{K+\rho}{K+1}\right)-\frac{K}{2}H\left(\frac{K-\rho}{K+1}\right)\,,
\end{align*}
where we used that $H(b) = H(\frac{1+\mu}{2}) = H(\frac{K}{K+1})$.
Note that $M_{1}(\rho)=1-H(\frac{1+\rho}{2})=1-H(\alpha).$ We want to show $M_{K}(\rho)\le M_{1}(\rho)$ for $K\ge1$.

Observe that $M_{K}(0)=0$ for all $K$, so it suffices to show that the derivative with respect to $\rho$ satisfies $M_{K}^{\prime}(\rho)\le M_{1}^{\prime}(\rho)$.

The derivative is calculated as:
\[
M_{K}^{\prime}(\rho)=\frac{K}{2(K+1)\ln 2}\ln\left(\frac{(1+\rho)(K+\rho)}{(1-\rho)(K-\rho)}\right).
\]
Let $A=\frac{1+\rho}{1-\rho}$. Then $M_{1}^{\prime}(\rho)=\frac{1}{2\ln 2}\ln(A)$. The inequality $M_{K}^{\prime}(\rho)\le M_{1}^{\prime}(\rho)$ is equivalent to:
\[
\frac{K}{K+1}\ln\left(A\frac{K+\rho}{K-\rho}\right)\le \ln A.
\]
Rearranging this inequality yields:
\[
K \ln\left(\frac{K+\rho}{K-\rho}\right)\le \ln A \Leftrightarrow K \ln\left(\frac{1+\rho/K}{1-\rho/K}\right)\le \ln\left(\frac{1+\rho}{1-\rho}\right).
\]
Let $f(x)=\frac{1}{x}\ln(\frac{1+x}{1-x})$. The inequality is equivalent to $f(\rho/K)\le f(\rho)$. The Taylor series is $f(x)=2\sum_{j=0}^{\infty}\frac{x^{2j}}{2j+1}$, which is strictly increasing for $x>0$. Since $K\ge1$, we have $\rho/K\le\rho$, thus $f(\rho/K)\le f(\rho)$.

The inequality $M_{K}^{\prime}(\rho)\le M_{1}^{\prime}(\rho)$ holds. Integrating from $\rho=0$ yields $M_{K}(\rho)\le M_{1}(\rho)=1-H(\alpha)$.

\bigskip
\noindent\textbf{Step 4: Bounding the Maximum (Case 2: $n<K$).}
If $n<K$, then $nC^{2}< R^{2}$. The Parseval constraint $\sum w_{i}\le R^{2}$ is inactive. We maximize the convex function $\sum\psi_{\mu}(w_{i})$ subject to the box constraints $0\le w_{i}\le C^{2}$. Therefore the maximum is attained at an extreme point. In addition, $\psi_\mu(C^2) = g_\mu(C)\ge 0$ (as it represents mutual information). Also, $\psi_\mu(0) = 0$, and so $\psi_\mu(C^2)\ge \psi_\mu(0)$. Therefore, the maximum is attained when $w_{i}=C^{2}$ for all $i\in[n]$.

The objective value is:
\[
\sum_{i=1}^n \psi_{\mu}(w_i)= n\psi_{\mu}(C^{2})=ng_{\mu}(C)<Kg_{\mu}(C)=M_{K}(\rho),
\]
Since $g_{\mu}(C)\ge0$ (as it represents mutual information) and $n<K$. From Step 3, we established $M_{K}(\rho)\le1-H(\alpha)$. 

In both cases, the total mutual information is bounded by $1-H(\alpha)$.

\section{Optimal Entropy Bound in the High Noise Regime}\label{sec:high_noise}

In this section, we focus on the main Courtade-Kumar conjecture in the high noise regime and prove Theorem \ref{thm:extended_range}.
\subsection{Proof of Theorem~\ref{thm:extended_range}}
The proof strategy follows the established path in \cite{samorodnitsky2016entropy}, relying on the combination of Fourier concentration and structural theorems for Boolean functions.

Theorem \ref{thm:linear_concentration_intro} implies that a highly informative function $f$ has most of its Fourier mass on level 1. Define $\xi:= \sum_{|S|\geq 2} \hat{f}(S)^2$. 
Theorem 5.5 in \cite{samorodnitsky2016entropy} at high level states that such a function must be close to a dictatorship. Concretely, by this theorem we have
\begin{align}
|\E[f]| &= O(\xi \sqrt{\ln(1/\xi)}),\label{eq:xi1}\\
\hat{f}(\{k\})^2 &\ge 1- \xi - O(\xi^2\ln(1/\xi))\,\label{eq:xi2}
\end{align}
for some $k\in[n]$.\footnote{Note that in our notation $f$ takes its values in $\{-1,1\}$ which correspond to $g$ notation in~\cite{samorodnitsky2016entropy}.}
The next step  involves analyzing the mutual information of near-dictatorship functions (Theorem~1.14 in~\cite{samorodnitsky2016entropy}), by which
 Conjecture~\ref{conj:CK} holds provided that
\begin{align}\label{eq:constraaint1}
    \xi\sqrt{\ln(1/\xi)}\le c_0\,,
\end{align}
for a small absolute constant $c_0$.

There is also another constraint that needs to be satisfied and is implicit in the proof of Theorem 1.14~\cite{samorodnitsky2016entropy}, which we discuss next.
Define $\alpha, \beta \ge0 $ as follows: $\beta = -(1/2) \hat{f}(\emptyset) = -(1/2)\E[f]$ and $\hat{f}(\{k\}) = (1-\alpha) (1-2\beta)$.  Also set $\gamma:=\alpha+\beta$. Then, another condition used in Proof of ~\cite[Theorem 1.14 (page 30)]{samorodnitsky2016entropy} is that 
\begin{align}\label{eq:constraint2}
\lambda+ \gamma\ln(1/\gamma) \le c_1
\end{align}
for some small absolute constant $c_1>0$. Invoking~\eqref{eq:xi1} and~\eqref{eq:xi2}, $\gamma = O(\xi\sqrt{\ln(1/\xi)})$. In Theorem~\ref{thm:linear_concentration_intro}, we proved that $\xi = O(\lambda)$ and so $\gamma = O(\lambda\sqrt{\ln(1/\lambda)})$.  Since $\ln(1/\gamma) = O(\ln(1/\lambda))$, we have:
$$ \gamma \ln(1/\gamma) = O(\lambda (\ln(1/\lambda))^{3/2}). $$

% The closeness parameter $\gamma$ (measuring the $L_2$ distance to the closest dictatorship) satisfies:
% $$ \gamma = O\left( \left(\sum_{|S|\geq 2} \hat{f}(S)^2\right) \sqrt{\ln(1/\lambda)} \right). $$

% Using our linear concentration bound $O(\lambda)$:
% $$ \gamma = O(\lambda \sqrt{\ln(1/\lambda)}). $$

% The final step involves analyzing the mutual information of near-dictatorships (Theorem 1.14 in \cite{samorodnitsky2016entropy}). This analysis confirms that the dictatorship is the maximizer, provided the noise is sufficiently high, specifically requiring the condition:
% $$ \gamma \ln(1/\gamma)+\lambda \leq c_0 $$
% for a small absolute constant $c_0$.

% We analyze the dominant term $\gamma \ln(1/\gamma)$. Since $\ln(1/\gamma) = O(\ln(1/\lambda))$, we have:
% $$ \gamma \ln(1/\gamma) = O(\lambda (\ln(1/\lambda))^{3/2}). $$

Hence, condition~\eqref{eq:constraint2} becomes 
\begin{align}\label{eq:constraint2-2}
\lambda+\lambda(\ln(1/\lambda))^{3/2}  \leq c'_1
\end{align}
for some absolute constant $c'_1>0$. Furthermore, using  that $\xi = O(\lambda)$, condition~\eqref{eq:constraaint1} is equivalent to 
\begin{align}\label{eq:constraint1-2}
\lambda \sqrt{\ln(1/\lambda)}\le c_0\,.
\end{align}
Condition~\eqref{eq:constraint2-2} already implies Condition \eqref{eq:constraint1-2}. In summary, Condition~\eqref{eq:constraint2-2} defines the threshold $\delta_{opt}$.
% $C'\lambda(\ln(1/\lambda))^{3/2} + \lambda \leq c_0$ for some constant $C'$. This inequality defines the threshold $\delta_{opt}$.

In contrast, the analysis in \cite{samorodnitsky2016entropy}, based on the $O(\lambda^{4/3})$ error bound, yielded a Fourier concentration of $O(\lambda^{1/3})$. This led to the bound $\gamma = O(\lambda^{1/3} \sqrt{\ln(1/\lambda)})$ and a condition dominated by $O(\lambda^{1/3}(\ln(1/\lambda))^{3/2})$.

Since $1 > 1/3$, the function $x(\ln(1/x))^{3/2}$ approaches zero significantly faster than $x^{1/3}(\ln(1/x))^{3/2}$ as $x \to 0$. Therefore, the threshold $\delta_{opt}$ satisfying the new, tighter inequality is strictly larger than the threshold derived from the $O(\lambda^{4/3})$ analysis.

\subsection{Proof of Theorem~\ref{thm:linear_concentration_intro}}\label{sec:implications}

The optimal error bound established in the previous section has significant consequences for the structural properties of highly informative functions which will be used next to prove Theorem~\ref{thm:linear_concentration_intro}.

% \subsection{Linear Fourier Concentration}

% The $O(\lambda^2)$ error bound allows us to derive the strongest possible Fourier concentration result within this asymptotic framework.

% \begin{theorem}[Linear Fourier Concentration]
% Let $f$ be a Boolean function satisfying $c \leq \E f \leq 1/2$ (for some absolute constant $c > 0$). If $I(f(X); Y) \geq 1 - H(\alpha)$, then
% $$ \sum_{|S|\geq 2} \hat{f}(S)^2 = O(\lambda). $$
% \end{theorem}

% \begin{proof}
Let $p = \E f$. We utilize the connection: $I(f(X); Y) = \Ent(T_\alpha f) + \Ent(T_\alpha(1-f))$.

We apply Theorem \ref{thm:optimal_entropy_intro} by normalizing $f$. Using $\Ent(h) = \E[h] \cdot \Ent(h/\E[h])$, we have $\Ent(T_\alpha f) = p \cdot \Ent(T_\alpha(f/p))$. Applying the theorem to $f/p$:
\begin{align*}
\Ent(T_\alpha f) &\leq p \left(\frac{L_1(f/p)}{2 \ln 2} \lambda + O(\lambda^2)\right) \\
&= \frac{L_1(f)}{p(2 \ln 2)}\lambda + O(\lambda^2).
\end{align*}
The $O(\lambda^2)$ constant remains controlled because $p$ is bounded away from 0 by $c$. Applying the same logic to $1-f$, and noting $L_1(f) = L_1(1-f)$:
\begin{align*}
I(f(X); Y) &\leq \left(\frac{1}{p} + \frac{1}{1-p}\right) \frac{L_1(f)}{2 \ln 2} \lambda + O(\lambda^2) \\
&= \frac{1}{p(1-p)} \frac{L_1(f)}{2 \ln 2} \lambda + O(\lambda^2).
\end{align*}

We use the Taylor expansion of the channel capacity: $1 - H(\alpha) = \frac{\lambda}{2 \ln 2} + O(\lambda^2)$. Combining this with the assumption $I(f(X); Y) \geq 1 - H(\alpha)$:
$$ \frac{\lambda}{2 \ln 2} + O(\lambda^2) \leq \frac{1}{p(1-p)} \frac{L_1(f)}{2 \ln 2} \lambda + O(\lambda^2). $$

Dividing by $\lambda/(2 \ln 2)$:
$$ 1 + O(\lambda) \leq \frac{L_1(f)}{p(1-p)} + O(\lambda). $$
Rearranging:
$$ p(1-p) - O(\lambda) \leq L_1(f). $$

By Parseval's identity, $\sum_{S\neq\emptyset} \hat{f}(S)^2 = p(1-p)$. Therefore, the Fourier weight on levels 2 and higher is:
$$ \sum_{|S|\geq 2} \hat{f}(S)^2 = p(1-p) - L_1(f) \leq O(\lambda). $$
%\end{proof}

This linear concentration is a significant improvement over the $O(\lambda^{1/3})$ bound established in \cite{samorodnitsky2016entropy}.

\section{Potential Applications}\label{sec:applications}

The results presented here enhance our understanding of information propagation in Boolean circuits and have several potential applications.

\subsection{Coding Theory and Polarization}

The Courtade-Kumar conjecture is intrinsically linked to the phenomenon of channel polarization, the basis for capacity-achieving polar codes. 

% The conjecture implies that under the polarization transform, channels tend towards being completely noisy or completely noiseless. Our results strengthen the theoretical foundation for why polarization occurs, particularly in the high noise regime. 

The polarization transforms a set of noisy channels towards either being completely noisy or completely noiseless~\cite{arikan2009channel}. The Courtade-Kumar conjecture implies that the resulting ``good'' channels are optimally represented by dictatorship functions, confirming that the polarization transform is moving in the right direction. Our results (Theorem~\ref{thm:extended_range}) strengthen the theoretical understanding of why this dictatorial structure remains the most stable ``sink'' for information, particularly in the high noise regime. 

The generalized bound (Theorem \ref{thm:generalized_thm1_intro}) may also find applications in analyzing polarization over asymmetric or biased channels.

\subsection{Hardness of Approximation and Dictatorship Testing}

In complexity theory, the analysis of Boolean functions under noise is fundamental to constructing Probabilistically Checkable Proofs (PCPs) and proving hardness of approximation results. Dictatorship tests often rely on the noise stability of the function, which is closely related to the mutual information studied here~\cite{parnas2002testing,fischer2004testing}.

Theorem \ref{thm:linear_concentration_intro} provides a robust structural characterization: functions that preserve significant information under high noise must be very close to dictatorships. The linear concentration bound $O(\lambda)$ provides a tighter quantitative relationship between information preservation and structure. This could potentially lead to more efficient dictatorship tests or tighter inapproximability gaps.

\subsection{Noise Sensitivity and Percolation}

The results also relate to the study of noise sensitivity and influences of Boolean functions. 
The influence of the $i$-th bit on a Boolean function $f$ is defined as $I_i (f) = \E[(f(x) - f(x\oplus e_i))^2]$ where $x \oplus e_i$ is the same vector but with the $i$-th bit flipped. The total influence is also defined as $I(f) = \sum_i I_i(f)$, and as shown in~\cite{li2020boolean} can be written in terms of its Fourier expansion as $I(f) = 4 \sum_{S\subseteq [n]} |S| \hat{f}(S)^2$. Our proof techniques for Theorem \ref{thm:linear_concentration_intro} can potentially be used to obtain tighter concentration bounds on total influence function. 
%The generalization to biased functions allows for analyzing a broader class of functions that arise naturally in percolation theory and random graph models.

\section{Conclusion}

We have made significant progress on two fronts related to the Courtade-Kumar conjecture. By resolving Open Question 1 of \cite{courtade2014boolean}, we established that the bound on the sum of coordinate-wise mutual information holds universally. Furthermore, by deriving an optimal $O(\lambda^2)$ error term for the entropy expansion in the high noise regime, we obtained a sharp linear Fourier concentration result. This refinement significantly extends the proven range of the conjecture, reinforcing the understanding that dictatorships are the most resilient functions against high levels of noise.

\section*{Acknowledgment}
The results in this paper were obtained with significant interaction with a larger version of Google's Deep Think Gemini-based model. The authors verified the entire paper and take full responsibility for all results. We thank Vincent Cohen-Addad, Lalit Jain, Jieming Mao, Vahab Mirrokni, and Song Zuo for their work on advanced Gemini models. We also thank the entire Deep Think team.

\bibliographystyle{acm}
\bibliography{Refs}
% \bibitem{JOW} S. Janson, K. Oleszkiewicz, and A. Woel, "Inﬂuence and stability for the product measure on the discrete cube," Manuscript, 2003.

%=====================
\clearpage
\section*{Appendix}
\subsection{Proof of Lemma~\ref{lem:fourier-prop}}
We first prove that $x_i(\tilde{b}) = x_i(b)$ for $i \neq j$.

Recall the definition of the Fourier coefficient $z_i(b) = \E [b(X^n) X_i]$.
Since $i\neq j$, $X_i$ is independent of $X_j$.
Let $X_{\sim j}$ denote the bits of $X^n$ excluding index $j$. We write $$z_i(b) = \mathbb{E}_{X_{\sim j}} \left[ \frac{X_i}{2} \left( b(X_{\sim j}, -1) + b(X_{\sim j}, 1) \right) \right]$$

Now consider the compressed function $\tilde{b}$. By definition, compression locally rearranges the pair $(b(x_{\sim j}, -1), b(x_{\sim j}, 1))$ into $(\tilde{b}(x_{\sim j}, -1), \tilde{b}(x_{\sim j}, 1))$ such that the sum is preserved :$$b(x_{\sim j}, -1) + b(x_{\sim j}, 1) = \tilde{b}(x_{\sim j}, -1) + \tilde{b}(x_{\sim j}, 1)$$

Substituting this back into the expectation:$$\begin{aligned}
z_i(b) &= \mathbb{E}_{X_{\sim j}} \left[ \frac{X_i}{2} \left( \tilde{b}(X_{\sim j}, -1) + \tilde{b}(X_{\sim j}, 1) \right) \right] \\
&= \mathbb{E} [\tilde{b}(X^n) X_i] = z_i(\tilde{b})
\end{aligned}$$
Thus, $z_i(\tilde{b}) = z_i(b)$ for $i \neq j$.

We next show that $|z_j(\tilde{b})| \ge |z_j(b)|$. Expand the Fourier coefficient for index $j$ by conditioning on $X_{\sim j}$:
$$z_j(b) = \frac{1}{2} \mathbb{E}_{X_{\sim j}} \left[ b(X_{\sim j}, 1) - b(X_{\sim j}, -1) \right]$$
By definition, compression locally rearranges the pair $(b(x_{\sim j}, -1), b(x_{\sim j}, 1))$ into $(\tilde{b}(x_{\sim j}, -1), \tilde{b}(x_{\sim j}, 1))$. Therefore, $|b(x_{\sim j}, 1) - b(x_{\sim j}, -1)| = \tilde{b}(x_{\sim j}, 1) - \tilde{b}(x_{\sim j}, -1)$ for any value of $x_{\sim j}$. Hence,
\begin{align*}
|z_j(b)|& = \frac{1}{2} \left|\mathbb{E}_{X_{\sim j}} \left[ b(X_{\sim j}, 1) - b(X_{\sim j}, -1) \right]\right|\\
&\le  \frac{1}{2} \mathbb{E}_{X_{\sim j}} \left[ |b(X_{\sim j}, 1) - b(X_{\sim j}, -1) |\right]\\
&= \frac{1}{2} \mathbb{E}_{X_{\sim j}} \left[ |\tilde{b}(X_{\sim j}, 1) - \tilde{b}(X_{\sim j}, -1) |\right]\\
&= \frac{1}{2} \mathbb{E}_{X_{\sim j}} \left[ \tilde{b}(X_{\sim j}, 1) - \tilde{b}(X_{\sim j}, -1) \right] = z_j(\tilde{b})\,,
\end{align*}
where in the first step we used the convexity of absolute value ($|\E[Y]|\le \E[|Y|]$). In the third equality, we used the assumption  that  $\tilde{b}$ is non-decreasing in the $j$-th bit. Note that this also implies that $z_j(\tilde{b})\ge 0$, which completes the proof. 

\bigskip

%=====================
\subsection{Proof of Lemma~\ref{lem:xi-pos}}
We can expand the expectation by conditioning on the value of the random variable $X_i$. Since $X$ is uniform, $X_i$ takes values $-1$ and $+1$ with probability $1/2$ each.

\begin{equation}
\begin{aligned}
z_i(b) &= \mathbb{E}[b(X) X_i] \\
&= \frac{1}{2} \mathbb{E}[b(X) X_i \mid X_i = -1] + \frac{1}{2} \mathbb{E}[b(X) X_i \mid X_i = 1]
\end{aligned}
\end{equation}
Substitute the value of $X_i$ into the expression:
\begin{equation}
\begin{aligned}
z_i(b) &= \frac{1}{2} \mathbb{E}[b(X_{\sim i}, -1) \cdot (-1)] + \frac{1}{2} \mathbb{E}[b(X_{\sim i}, 1) \cdot (1)] \\
&= \frac{1}{2} \mathbb{E}_{x_{\sim i}} \left[ b(x_{\sim i}, 1) - b(x_{\sim i}, -1) \right]
\end{aligned}
\end{equation}

    We are given that $b$ is non-decreasing bit-wise. Therefore, for every possible setting of the other bits $x_{\sim i}$, we have
$b(x_{\sim i}, 1) \ge b(x_{\sim i}, -1)$.
    Thus, $z_i(b)\ge 0$.
%====================
\subsection{Proof of Lemma~\ref{lem:boolean-prop}}
Let $E_+ = \mathbb{E}[b(X) \mid X_i = +1]$ and $E_- = \mathbb{E}[b(X^n) \mid X_i = -1]$
receptively denote the average value of the function when the $i$-th bit is fixed to $+1$ and $-1$.
Since $b$ outputs values in $\{-1, 1\}$, we have $-1 \le E_+, E_- \le 1$. In addition, we have
\begin{align*}
\mu &= \frac{1}{2}E_+ + \frac{1}{2}E_-\\
z_i &= \mathbb{E}[b(X)X_i] = \frac{1}{2}E_+ \cdot 1 + \frac{1}{2}E_- \cdot (-1) = \frac{E_+ - E_-}{2}
\end{align*}
Hence $\mu+z_i = E_+$ and $\mu-z_i  = E_-$ and the claim follows from the fact that $E_+\le 1$ and $E_-\ge -1$.

%================================
\bigskip

\section{Proof of Theorem~\ref{thm:optimal_entropy_intro}}
\medskip

$\bullet$ {\bf Preliminaries and Notation.} We analyze functions on the Boolean hypercube $\{-1, 1\}^n$. We follow the setup used in \cite{samorodnitsky2016entropy}. Let $f: \{-1, 1\}^n \to \R$ be a bounded, non-negative function, normalized such that $\E[f] = 1$. We assume $\|f\|_\infty \leq M$. (When analyzing a Boolean function $b$, we typically look at the indicator function of the event $b(x)=1$).

Let $\alpha \in [0, 1/2]$ be the noise parameter (we use $\alpha$ here instead of $\epsilon$ to maintain consistency with Section II, although $\epsilon$ is common in this literature). The noise operator $T_\alpha$ acts on $f$ as $(T_\alpha f)(x) = \E_y[f(y)]$, where $y$ is obtained from $x$ by flipping coordinates with probability $\alpha$. Let $\rho = 1-2\alpha$. In the Fourier domain, $T_\alpha$ acts as a multiplier:
$$ \widehat{T_\alpha f}(S) = \rho^{|S|} \hat{f}(S). $$
We define the noise parameter $\lambda = \rho^2 = (1-2\alpha)^2$. We consider the high noise regime, where $\lambda$ is small.

We decompose $f$ into its even part $f_0$ and odd part $f1$:
$$ f_0(x) = \frac{f(x) + f(-x)}{2}, \quad f_1(x) = \frac{f(x) - f(-x)}{2}. $$
Note that $\E[f_0] = 1$.

We define the noisy versions:
$$ F = T_\alpha f_0, \quad Z = T_\alpha f_1. $$
Note that $T_\alpha f = F + Z$. We define $V = F - 1$. Since $\E[F] = 1$, we have $\E[V] = 0$.

The Fourier coefficients satisfy $\hat{F}(S) = \lambda^{|S|/2} \hat{f_0}(S)$ and $\hat{Z}(S) = \lambda^{|S|/2} \hat{f_1}(S)$. $F$ (and $V$) are supported on even Fourier levels, while $Z$ is supported on odd levels.

The level 1 Fourier weight is $L_1(f) = L_1(f_1)$.

We utilize the entropy decomposition established in \cite{samorodnitsky2016entropy} (Lemma 6.2):

\begin{lemma}[Entropy Decomposition]\label{lem:entropy_decomp}
$$ \Ent(T_\alpha f) \leq \Ent(F) + \frac{1}{2 \ln 2} \E\left[\frac{Z^2}{F}\right] + O\left(\E\left[\frac{Z^4}{F^3}\right]\right). $$
\end{lemma}

It is known that $\Ent(F) = O(\lambda^2)$ (Lemma 5.4 in \cite{samorodnitsky2016entropy}).

We also rely on a crucial dominance property which is proven below. 
\begin{lemma}[Dominance]\label{lem:dominance}
$F \geq 0$ and $|Z| \leq F$.
\end{lemma}

\medskip

$\bullet$ {\bf Hypercontractivity and Moment Bounds}
\\\\
We also rely on the Bonami-Beckner Hypercontractivity Theorem. A standard application is for homogeneous polynomials.

\begin{corollary}[Hypercontractivity for Homogeneous Polynomials]\label{cor:hyper_homog}
Let $h_k$ be a homogeneous polynomial of degree $k$. For $q \geq 2$,
$$ \|h_k\|_q \leq (\sqrt{q-1})^k \|h_k\|_2. $$
\end{corollary}

We must be careful when applying this to functions supported on multiple levels. We use the Minkowski inequality combined with Corollary \ref{cor:hyper_homog} to establish rigorous moment bounds. We prove the following lemma below.

\begin{lemma}[Moment Bounds]\label{lem:moment_bounds}
Assuming $f$ is bounded by $M$, the following bounds hold as $\lambda \to 0$:
\begin{enumerate}
    \item $\E[V^2] = O(\lambda^2)$.
    \item $\E[|V|^3] = O(\lambda^3)$.
    \item $\E[Z^2] = \lambda L_1(f) + O(\lambda^3)$.
    \item $\E[Z^4] = O(\lambda^2)$.
    \item $\E[Z^2V] = O(\lambda^2)$.
\end{enumerate}
\end{lemma}

We now bound the higher order term in the entropy decomposition (Lemma \ref{lem:entropy_decomp}). We prove the following lemma below.

\begin{lemma}[Higher Order Term]\label{lem:higher_order}
$\E[Z^4/F^3] = O(\lambda^2)$.
\end{lemma}

\medskip

$\bullet$ {\bf Optimal Asymptotic Entropy Bound.} Given the above supporting lemmas, which we prove below, we now prove the main technical result of this section (Theorem~\ref{thm:optimal_entropy_intro}), achieving the optimal $O(\lambda^2)$ error bound by employing a direct Taylor expansion.

% \begin{theorem}[Optimal Asymptotic Entropy Bound]
% For any bounded nonnegative non-zero function $f$ with $\E[f] = 1$:
% $$ \Ent(T_\alpha f) \leq \left(\frac{1}{2 \ln 2}L_1(f)\right) \cdot \lambda + O(\lambda^2). $$
% \end{theorem}

%\begin{proof}
We start from the entropy decomposition (Lemma \ref{lem:entropy_decomp}):
$$ \Ent(T_\alpha f) \leq \Ent(F) + \frac{1}{2 \ln 2} \E\left[\frac{Z^2}{F}\right] + O\left(\E\left[\frac{Z^4}{F^3}\right]\right). $$
We know $\Ent(F) = O(\lambda^2)$ and by Lemma \ref{lem:higher_order}, the error term is $O(\lambda^2)$. We analyze $B = \E[Z^2/F]$.

We use the Taylor expansion of $1/F = 1/(1+V)$ around $V=0$ with the exact remainder:
$$ \frac{1}{1+V} = 1 - V + \frac{V^2}{1+V}. $$
This holds when $F = 1+V > 0$. Let $S_+ = \{x : F(x) > 0\}$.

$$ B = \E\left[\frac{Z^2}{F} \mathbb{I}{S+}\right] = \E\left[Z^2 \left(1 - V + \frac{V^2}{F}\right) \mathbb{I}{S+}\right]. $$

\textbf{Term 1 (Expansion):} $E_T = \E[Z^2(1-V)\mathbb{I}{S+}]$.
By Lemma \ref{lem:dominance}, if $F(x)=0$, then $Z(x)=0$. Therefore, the indicator is redundant and we have
$$ E_T = \E[Z^2] - \E[Z^2V]. $$
By Lemma \ref{lem:moment_bounds}(3) and (5):
\begin{align*}
E_T &= (\lambda L_1(f) + O(\lambda^3)) - O(\lambda^2) = \lambda L_1(f) + O(\lambda^2).
\end{align*}

\textbf{Term 2 (Remainder):} $R_T = \E\left[\frac{Z^2V^2}{F} \mathbb{I}{S+}\right]$.
Using dominance $|Z| \leq F$, $Z^2/F \leq F$.
$$ R_T \leq \E[F V^2 \mathbb{I}{S+}] \leq \E[F V^2]. $$
$$ \E[F V^2] = \E[(1+V)V^2] = \E[V^2] + \E[V^3]. $$
By Lemma \ref{lem:moment_bounds}(1) and (2), $R_T = O(\lambda^2) + O(\lambda^3) = O(\lambda^2)$.

Combining the terms for $B$:
$$ B = \lambda L_1(f) + O(\lambda^2). $$

Substituting this back into the entropy decomposition:
\begin{align*}
\Ent(T_\alpha f) &\leq O(\lambda^2) + \frac{1}{2 \ln 2}(\lambda L_1(f) + O(\lambda^2)) + O(\lambda^2) \\
&= \frac{L_1(f)}{2 \ln 2} \lambda + O(\lambda^2).
\end{align*}

We next show that the $O(\lambda^2)$ error term is asymptotically tight. We demonstrate this by analyzing the the dictatorship. functions which are believed to maximize the entropy (and mutual information).

Consider the Boolean dictatorship function $f(x)=x_1$. The mutual information is exactly the channel capacity, $I(f(X); Y) = 1 - H(\alpha)$, where $H(\alpha)$ is the binary entropy function (in bits). We analyze the Taylor expansion of the capacity around $\alpha=1/2$ (where $\rho=0$).

Recall $\rho = 1-2\alpha$ and $\lambda = \rho^2$. The expansion of the binary entropy function $H(\alpha)$ around $\alpha=1/2$ yields
$$
H(\alpha) = 1 - \frac{1}{2\ln 2}\rho^2 - \frac{1}{12\ln 2}\rho^4 + O(\rho^6).
$$
Therefore, the capacity is:
$$
1 - H(\alpha) = \frac{1}{2\ln 2}\lambda + \frac{1}{12\ln 2}\lambda^2 + O(\lambda^3).
$$
For the dictatorship function, $L_1(f)=1$. Thus, the expansion matches the form in the theorem statement:
$$
I(f(X); Y) = \left(\frac{L_1(f)}{2\ln 2}\right)\lambda + \Theta(\lambda^2).
$$
Since the expansion for the maximizing function includes a $\Theta(\lambda^2)$ term, the general upper bound established in the theorem statement cannot be asymptotically improved beyond $O(\lambda^2)$.
%\end{proof}
%=================
\bigskip

\subsection{Proof of Lemma~\ref{lem:dominance}}
Note that by definition of the even and odd parts of $f$, we have
\begin{align*}
f_0(x)+f_1(x) &= 2f(x)\\
f_0(x)-f_1(x) &= 2f(-x)
\end{align*}
Since $f \geq 0$, we have $|f_1(x)| \leq f_0(x)$. The noise operator $T_\alpha$ is a positive operator, preserving non-negativity and dominance. Thus $|Z| = |T_\alpha f_1| \leq T_\alpha |f_1| \leq T_\alpha f_0 = F$.

%============
\bigskip

\subsection{Proof of Lemma~\ref{lem:moment_bounds}}
We assume $\lambda$ is small enough (e.g., $3\lambda < 1$). Since $f$ is bounded, its $L_2$ norm is also bounded ($\|f\|_2 \leq M$).

\textbf{Part 1: $\E[V^2] = O(\lambda^2)$.}
$V$ is supported on even levels $\geq 2$.
\begin{align*}
\E[V^2] &= \sum_{k\geq 2, \text{even}} \lambda^k \sum_{|S|=k} \hat{f_0}(S)^2 \\
&\leq \lambda^2 \sum_{S \neq \emptyset} \hat{f_0}(S)^2 = \lambda^2 \Var(f_0) = O(\lambda^2).
\end{align*}

\textbf{Part 3: $\E[Z^2] = \lambda L_1(f) + O(\lambda^3)$.}
$Z$ is supported on odd levels $\geq 1$.
\begin{align*}
\E[Z^2] &= \lambda L_1(f_1) + \sum_{k\geq 3, \text{odd}} \lambda^k \sum_{|S|=k} \hat{f_1}(S)^2 \\
&= \lambda L_1(f) + O(\lambda^3).
\end{align*}

\textbf{Part 4: $\E[Z^4] = O(\lambda^2)$.}
We decompose $Z = Z_1 + Z_{\geq 3}$. We use the Minkowski inequality for the $L_4$ norm: $\|Z\|_4 \leq \|Z_1\|_4 + \|Z_{\geq 3}\|_4$.
\begin{enumerate}
\item Bounding $\|Z_1\|_4$. $Z_1$ is degree 1. Use Corollary \ref{cor:hyper_homog} ($q=4, k=1$).
$$ \|Z_1\|_4 \leq \sqrt{3} \|Z_1\|_2. $$
$\|Z_1\|_2^2 = O(\lambda)$. Thus, $\|Z_1\|_4 = O(\lambda^{1/2})$.

\item Bounding $\|Z_{\geq 3}\|_4$. Use Minkowski inequality on $Z_{\geq 3} = \sum_{k\geq 3, \text{odd}} Z_k$.
$$ \|Z_{\geq 3}\|_4 \leq \sum_{k\geq 3, \text{odd}} \|Z_k\|_4. $$
Applying hypercontractivity to each $Z_k$:
$$ \|Z_k\|_4 \leq (\sqrt{3})^k \|Z_k\|_2. $$
Since $\|\hat{f}_k\|_2 \leq M$, $\|Z_k\|_2 \leq M \lambda^{k/2}$.
$$ \|Z_k\|_4 \leq M (\sqrt{3\lambda})^k. $$
We sum this geometric series:
$$ \|Z_{\geq 3}\|_4 \leq M \sum_{k\geq 3, \text{odd}} (\sqrt{3\lambda})^k = M \frac{(\sqrt{3\lambda})^3}{1-3\lambda} = O(\lambda^{3/2}). $$

\item Combining the bounds.
$$ \|Z\|_4 \leq O(\lambda^{1/2}) + O(\lambda^{3/2}) = O(\lambda^{1/2}). $$
Therefore, $\E[Z^4] = \|Z\|_4^4 = O(\lambda^2)$.
\end{enumerate}

\textbf{Part 2: $\E[|V|^3] = O(\lambda^3)$.}
We analyze $V=V_2+V_{\geq 4}$ using the $L_3$ norm.

\begin{enumerate}
\item Bounding $\|V_2\|_3$. Use $q=3, k=2$. The constant is 2.
$$ \|V_2\|_3 \leq 2\|V_2\|_2. $$
$\|V_2\|_2^2 = O(\lambda^2)$. Thus, $\|V_2\|_3 = O(\lambda)$.

\item Bounding $\|V_{\geq 4}\|_3$.
$$ \|V_{\geq 4}\|_3 \leq \sum_{k\geq 4, \text{even}} \|V_k\|_3 \leq \sum_{k\geq 4, \text{even}} M (\sqrt{2\lambda})^k = O(\lambda^2). $$

\item Combining the bounds. $\|V\|_3 = O(\lambda)$. Therefore, $\E[|V|^3] = O(\lambda^3)$.
\end{enumerate}

\textbf{Part 5: $\E[Z^2V] = O(\lambda^2)$.}
We use the Cauchy-Schwarz inequality:
$$ |\E[Z^2V]| \leq \sqrt{\E[Z^4]\E[V^2]}. $$
Using Part 4 and Part 1:
$$ |\E[Z^2V]| \leq \sqrt{O(\lambda^2) \cdot O(\lambda^2)} = O(\lambda^2). $$
\bigskip

%================
\subsection{Proof of Lemma~\ref{lem:higher_order}}

We use a constant threshold $\Delta = 1/2$ to split the expectation.
$$ \E\left[\frac{Z^4}{F^3}\right] = \E\left[\frac{Z^4}{F^3} \mathbb{I}{F \leq 1/2}\right] + \E\left[\frac{Z^4}{F^3} \mathbb{I}{F > 1/2}\right]. $$
\medskip

\noindent\textbf{Term 2 (F is large):}
When $F > 1/2$, $1/F^3 < 8$.
$$ \E\left[\frac{Z^4}{F^3} \mathbb{I}_{F > 1/2}\right] \leq 8 \E[Z^4]. $$
By Lemma \ref{lem:moment_bounds}(4), $\E[Z^4] = O(\lambda^2)$. So Term 2 is $O(\lambda^2)$.
\medskip

\noindent\textbf{Term 1 (F is small):}
By Lemma \ref{lem:dominance}, $|Z| \leq F$. So $Z^4/F^3 \leq F$.
$$ \E\left[\frac{Z^4}{F^3} \mathbb{I}{F \leq 1/2}\right] \leq \E[F \mathbb{I}{F \leq 1/2}]. $$
Since $F \leq 1/2$ on this indicator,
$$ \E[F \mathbb{I}_{F \leq 1/2}] \leq \frac{1}{2} P(F \leq 1/2). $$
$P(F \leq 1/2) = P(V \leq -1/2)$.
By Chebyshev's inequality and Lemma \ref{lem:moment_bounds}(1):
$$ P(|V| \geq 1/2) \leq 4 \E[V^2] = O(\lambda^2). $$
So Term 1 is $O(\lambda^2)$.

\end{document}